\documentclass[twocolumn,showpacs,preprintnumbers,amsmath,amssymb]{revtex4}

\usepackage{graphicx}
\usepackage{dcolumn}
\usepackage{bm}
\usepackage{amsmath}

\begin{document}
\title{Information filtering based on transferring similarity}
\author{Duo Sun$^{1}$}
\author{Tao Zhou$^{1,2}$}
\email{zhutou@ustc.edu}
\author{Jian-Guo Liu$^{1,2}$}
\author{Run-Ran Liu$^{1}$}
\author{Chun-Xiao Jia$^{1}$}
\author{Bing-Hong Wang$^{1,3}$}

\affiliation{$^{1}$ Department of Modern Physics and Nonlinear
Science Center, University of Science and Technology of China, Hefei
Anhui, 230026, People's Republic of China\\
$^{2}$ Department of Physics, University of Fribourg, Chemin du
Musee 3, CH-1700 Fribourg, Switzerland\\
$^{3}$ Research Center for Complex System Science, University of
Shanghai for Science and Technology, Shanghai, 200093, People's
Republic of China}
\date{\today}

\begin{abstract}
In this Brief Report, we propose a new index of user similarity,
namely the transferring similarity, which involves all high-order
similarities between users. Accordingly, we design a modified
collaborative filtering algorithm, which provides remarkably higher
accurate predictions than the standard collaborative filtering. More
interestingly, we find that the algorithmic performance will
approach its optimal value when the parameter, contained in the
definition of transferring similarity, gets close to its critical
value, before which the series expansion of transferring similarity
is convergent and after which it is divergent. Our study is
complementary to the one reported in [E. A. Leicht, P. Holme, and M.
E. J. Newman, Phys. Rev. E {\bf 73} 026120 (2006)], and is relevant
to the missing link prediction problem.
\end{abstract}


\maketitle

With the exponential growth of the Internet \cite{NJP} and the
World-Wide-Web \cite{Broder00}, a prominent challenge for modern
society is the information overload. Since there are enormous data
and sources, people never have time and vigor to find out those most
relevant for them. A landmark for solving this problem is the use of
search engine \cite{Brin1998,Kleinberg1999}. However, a search
engine could only find the relevant web pages according to the input
keywords without taking into account the personalization, and thus
returns the same results regardless of users' habits and tastes.
Thus far, with the help of {\it Web2.0} techniques, personalized
recommendations become the most promising way to efficiently filter
out the information overload \cite{Adomavicius05}. Motivated by the
significance in economy and society, devising efficient and accurate
recommendation algorithms becomes a joint focus from theoretical
studies \cite{Adomavicius05} to e-commerce applications
\cite{Schafer2001}. Various kinds of algorithms have been proposed,
such as collaborative filtering (CF)
\cite{Herlocker2004,Konstan1997}, content-based methods
\cite{Balab97,Pazzani99}, spectral analysis
\cite{Billsus1998,Sarwar2000a,Jie2008}, principle component analysis
\cite{Goldberg2001}, network-based inference
\cite{Zhang2007a,Zhang2007b,Zhou2007,ZhouT2008}, and so on.

A recommender system consists of users and objects, and each user
has rated some objects. Denoting the user set as $U=\{u_1, u_2,
\cdots, u_N\}$ and the object set as $O=\{o_1, o_2, \cdots, o_M\}$,
the system can be fully described by an $N\times M$ rating matrix
$\textbf{V}$, with $v_{i\alpha}\neq 0$ denoting the rating user
$u_i$ gives to object $o_\alpha$. If $u_i$ has not yet evaluated
$o_\alpha$, $v_{i\alpha}$ is set as zero. CF system has been one of
the most successfully and widest used recommender systems since its
appearance in the mid-1990s \cite{Herlocker2004,Konstan1997}. Its
basic idea is that the user will be recommended objects based on the
weighted combination of similar users' opinions. In the standard CF,
the predicted rating $v'_{i\alpha}$ from user $u_i$ to object
$o_\alpha$ is set as:
\begin{equation}\label{Equ1}
v'_{i\alpha}=\bar{v}_{i}+I\negthickspace\sum_j\negthickspace{s_{ij}}(v_{j\alpha}-\bar{v}_{j}),
\end{equation}
where $s_{ij}$ is the similarity between $u_i$ and $u_j$,
$\bar{v}_{i}$ means the average rating of $u_i$ and
$I=(\sum_js_{ij})^{-1}$ serves as the normalization factor. Here,
$j$ runs over all users having rated object $o_\alpha$ excluding
$u_i$ himself. The similarity, $s_{ij}$, plays a crucial role in
determining the algorithmic accuracy. In the implementation, the
similarity between every pair of users is calculated firstly, and
then the predict ratings by Eq. (\ref{Equ1}). Various similarity
measures has been proposed, among which the Pearson correlation
coefficient is the widest used \cite{Herlocker2004}, as:
\begin{equation}
s_{ij}=\frac{\sum_{c}(v_{ic}-\bar{v}_i)(v_{jc}-\bar{v}_j)}{\sqrt{\sum_{\alpha}(v_{i\alpha}-\bar{v}_i)^2}
\sqrt{\sum_{\beta}(v_{j\beta}-\bar{v}_j)^2}},
\end{equation}
where $c$, $\alpha$ and $\beta$ run over all the objects commonly
selected by user $i$ and $j$. All diagonal elements in the
similarity matrix are set to be zero.

\begin{figure}
\scalebox{0.8}[0.8]{\includegraphics{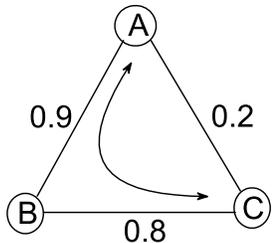}}
\caption{Illustration for transferring similarity.}\label{Fig.1}
\end{figure}

Several algorithms \cite{JGL2008,RL2008,JGL2009} have recently been
proposed to improve the accuracy of the standard CF via modifying
the definition of user-user similarity. However, all those
algorithms have not fully addressed the similarity induced by
indirect relationship, say, the high-order correlations. Note that,
the Pearson correlation coefficient, $s_{ij}$, considers only the
direct correlation. We argue that to appropriately measure the
similarities between users, the indirect correlations should also be
taken into consideration. To make our idea clearer, we draw an
illustration in Fig. \ref{Fig.1}. Suppose there are three users,
labeled as $A$, $B$ and $C$. Although the similarity between user
$A$ and $C$ is quite small, $A$ and $C$ are both very similar with
$B$. Actually, $A$, $B$ and $C$ may share very similar tastes, and
the very small similarity between $A$ and $C$ may be caused by the
sparsity of the data. That is to say, $A$ and $C$ has a very few
commonly selected objects. The sparsity of data set makes the direct
similarity less accurate, and thus we expect a new measure of
similarity properly integrating high-order correlations may perform
better.

\begin{figure}[b]
\scalebox{0.8}[0.8]{\includegraphics{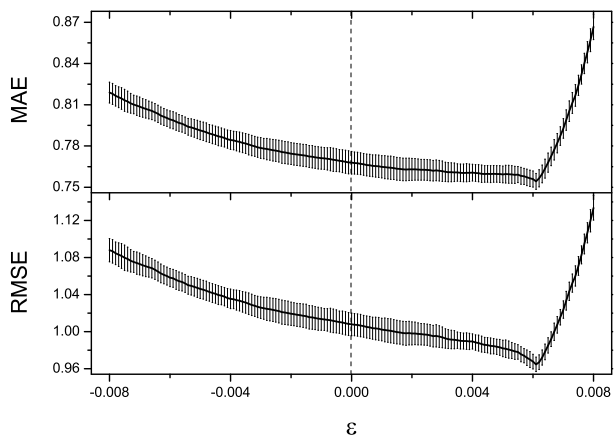}}
\caption{Prediction accuracy of the present algorithm, measured by
MAE and RSME, as functions of $\varepsilon$. The transferring
similarities are directly obtained by Eq. (5). The numerical results
are averaged over 20 independent runs, each corresponds to a random
division with training set containing about 90\% of data while the
probe consisted of the remain 10\%. The error bars denote the
standard deviations of the 20 samples.}\label{Fig2}
\end{figure}

\begin{figure}[b]
\scalebox{0.8}[0.8]{\includegraphics{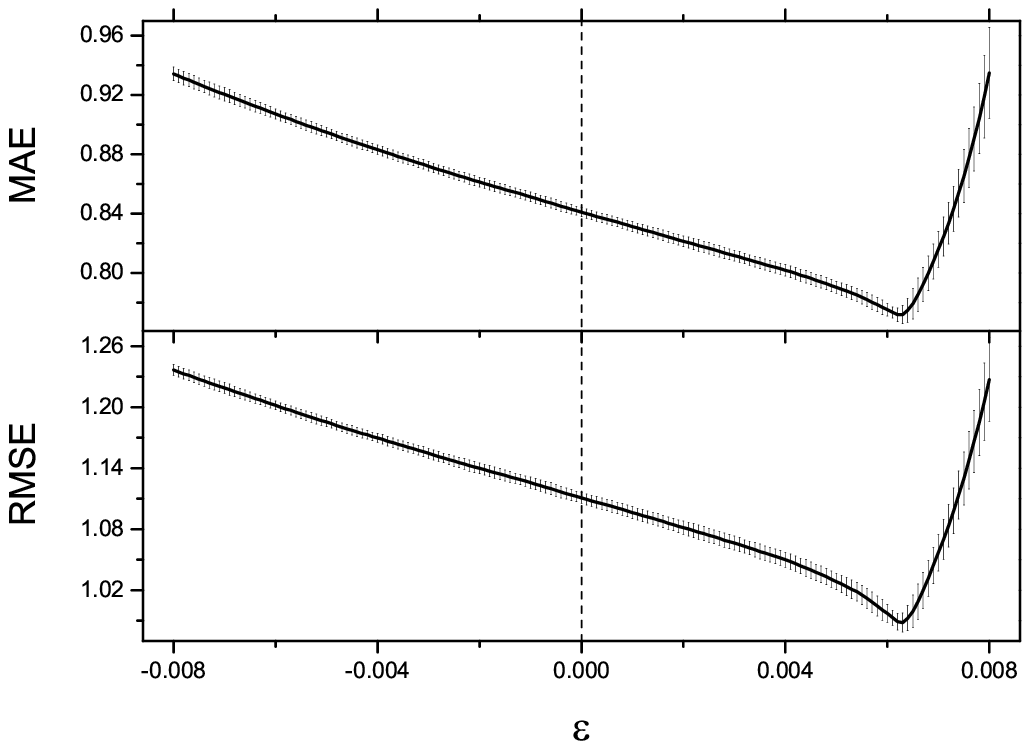}}
\caption{Prediction accuracy of the present algorithm, where the
division of training set and probe is 50\% vs. 50\%. Other
conditions are the same as what presented in Fig. 2.}\label{Fig3}
\end{figure}

\begin{figure}[b]
\scalebox{0.8}[0.8]{\includegraphics{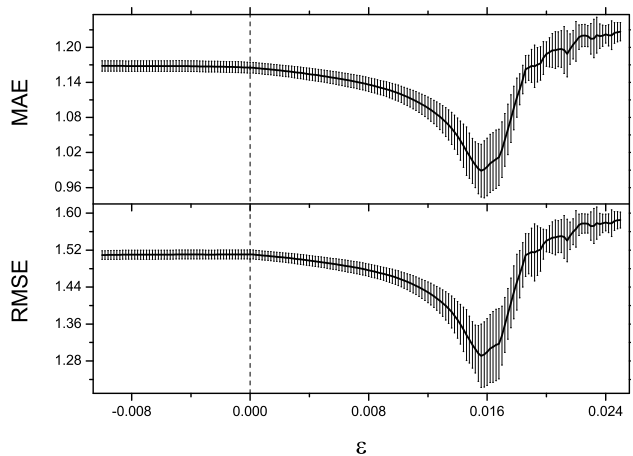}}
\caption{Prediction accuracy of the present algorithm, where the
division of training set and probe is 10\% vs. 90\%. Other
conditions are the same as what presented in Fig. 2.}\label{Fig4}
\end{figure}

Denoting $\varepsilon$ a decay factor of similarity transferred by a
medi-user, a self-consistent definition of \emph{transferring
similarity} can be written as:
\begin{equation}
t_{ij}=\varepsilon\sum_v s_{iv}t_{vj}+s_{ij},
\end{equation}
where $s_{ij}$ is the direct similarity as shown in Eq. (2). The
parameter $\varepsilon$ can be considered as the rate of information
aging by transferring one step further \cite{Stojmirovic2007}.
Clearly, the transferring similarity will degenerate to the
traditional Pearson correlation coefficient when $\varepsilon=0$.
Denoting $\textbf{S}=\{s_{ij}\}_{N\times N}$ and
$\textbf{T}=\{t_{ij}\}_{N\times N}$ the direct similarity matrix and
the transferring similarity matrix, Eq. (3) can be rewritten in a
matrix form, as:
\begin{equation}
\textbf{T}=\varepsilon\textbf{S}\textbf{T}+\textbf{S},
\end{equation}
whose solution is
\begin{equation}
\textbf{T}=(\textbf{1}-\varepsilon\textbf{S})^{-1}\textbf{S}.
\end{equation}
Accordingly, the prediction score reads
\begin{equation}
v'_{i\alpha}=\bar{v}_{i}+I'\negthickspace\sum_jt_{ij}(v_{j\alpha}-\bar{v}_{j}),
\end{equation}
where multiplier $I'=(\sum_jt_{ij})^{-1}$ serves as the normalizing
factor and $j$ runs over all users having rated object $o_\alpha$
excluding $u_i$ himself.

\begin{table}
\caption{The optimal and maximal values of $\varepsilon$ for the
three cases corresponding to Figs. 2-4. $\varepsilon_{\texttt{max}}$
is obtained by averaging 20 independent runs, and we have checked
that in each run $\varepsilon_{\texttt{opt}}$ is always a little bit
smaller than $\varepsilon_{\texttt{max}}$. The resolution of
$\varepsilon$ is $10^{-3}$ since for higher resolution (e.g.,
$10^{-4}$), the difference between two neighboring data point is
very small, and the optimal value is not distinguishable with the
presence of fluctuations.}
\begin{center}
\begin{tabular} {cccc}
  \hline \hline
   Data Divisions     & 90\% vs. 10\%  &  50\% vs. 50\%  &  10\% vs. 90\% \\
   \hline
$\varepsilon_{\texttt{opt}}$     & 0.0061  &  0.0063  &  0.0156 \\
$\varepsilon_{\texttt{max}}$     & 0.006136  &  0.006311  &  0.015642 \\
   \hline \hline
\end{tabular}
\end{center}
\end{table}

To test the algorithmic accuracy, we use a benchmark data set,
namely \emph{MovieLens}, which consists of $N=943$ users, $M=1682$
objects, and $10^5$ discrete ratings from 1 to 5. The sparsity of
the rating matrix $\textbf{V}$ is about 6\%. We first randomly
divide this data set into two parts: one is the training set,
treated as known information, and the other is the probe, whose
information is not allowed to be used for prediction. Then we make a
prediction for every entry contained in the probe (resetting
$v'_{i\alpha}=5$ and $v'_{i\alpha}=1$ in the case of
$v'_{i\alpha}>5$ and $v'_{i\alpha}<1$, respectively), and measure
the difference between the predicted rating $v'_{i\alpha}$ and the
actual rating $v_{i\alpha}$. For evaluating the accuracy of
recommendations, many different metrics have been proposed
\cite{Herlocker2004}. We choose two commonly used measures:
\emph{root-mean-square error} (RMSE) and \emph{mean absolute error}
(MAE). They are defined as
\begin{subequations}\label{subequation}
\begin{align}
{\rm RMSE}&=\sqrt{\sum_{(i,\alpha)}(v'_{i\alpha}-v_{i\alpha})^{2}/E}, \label{sub1}\\
{\rm MAE}&=\frac{1}{E}\sum_{(i,\alpha)}|v'_{i\alpha}-v_{i\alpha}|,
\end{align}
\end{subequations}
where the subscript $(i,\alpha)$ runs over all the elements in the
probe, and $E$ is the number of those elements.

In Figs. 2-4, we report the numerical results about the algorithmic
accuracy, where the divisions of training set and probe are 90\% vs.
10\%, 50\% vs. 50\%, and 10\% vs. 90\%, respectively. In every case,
there exists an optimal value of $\varepsilon$, denoted by
$\varepsilon_{\texttt{opt}}$, corresponding to both the lowest MAE
and the lowest RMSE. Around the optimal value,
$\varepsilon_{\texttt{opt}}$, the present algorithm obviously
outperforms the standard CF. The optimal values are different for
different cases, and the one corresponding to sparser data is
larger.

To get some insights about the physical meaning of
$\varepsilon_{\texttt{opt}}$, we expand Eq. (5) by a power series,
as:
\begin{equation}
\textbf{T}=\textbf{S}+\varepsilon\textbf{S}^{2}+\varepsilon^2\textbf{S}^{3}+\cdots.
\end{equation}
Note that, this formula is also of practical significance since to
directly inverse $(\textbf{1}-\varepsilon \textbf{S})$ takes long
time for huge-size systems, and the cutoff of Eq. (8),
\begin{equation}
\textbf{T}=\textbf{S}+\varepsilon\textbf{S}^{2}+\cdots+\varepsilon^n\textbf{S}^{n+1},
\end{equation}
is usually used as an approximation in the implementation (in this
paper, since the system size in not too large, we always directly
use Eq. (5) to obtain the transferring similarity matrix). However,
even if $(\textbf{1}-\varepsilon \textbf{S})$ is inversable, Eq. (8)
may not be convergent. Actually, Eq. (8) is convergent if and only
if all the eigenvalues of $(\textbf{1}-\varepsilon \textbf{S})$ are
strictly smaller than 1. The mathematical proof of a very similar
proposition using \emph{Jordan matrix decomposition} can be found in
Ref. \cite{Stojmirovic2007}. Although Ref. \cite{Stojmirovic2007}
only gives the proof of the sufficient condition, the necessary
condition can be proved in an analogical way. Accordingly, there
exists a critical point of $\varepsilon$, before which the spectral
radius of $\varepsilon \textbf{S}$ is less than 1 and after which it
exceeds 1. Since this critical value is also the maximal value of
$\varepsilon$ that keeps the convergence of Eq. (8), we denote it by
$\varepsilon_{\texttt{max}}$. The optimal and maximal values of
$\varepsilon$ for the three cases corresponding to Figs. 2-4 is
presented in Table 1. It is very interesting that
$\varepsilon_{\texttt{opt}}$ is always smaller yet very close to
$\varepsilon_{\texttt{max}}$.

In summary, we designed an improved collaborative filtering
algorithm based on a newly proposed similarity measure, namely the
transferring similarity. Different from the traditional definitions
of similarity that consider the direct correlation only, the
transferring similarity integrates all the high-order (i.e.,
indirect) correlations. The numerical testing on a benchmark data
set has demonstrated the improvement of algorithmic accuracy
compared with the standard CF algorithm. Very recently, Zhou {\it et
al.} \cite{Zhou2008b} and Liu {\it et al.} \cite{JGL2009} proposed
some modified recommendation algorithms under the frameworks of
collaborative filtering \cite{JGL2009} and random-walk-based
recommendations \cite{Zhou2008b},respectively. By taking into
account both the direct and the second order correlations, their
algorithms can remarkably enhance the prediction accuracy. These
work can be considered as a bridge connecting the
nearest-neighborhood-based information filtering algorithms and the
present work.

Very interestingly, we found that the optimal value of $\varepsilon$
is always smaller yet very close to the maximal value of
$\varepsilon$ that guarantees the convergence of power series
expansion of the transferring similarity. The significance of this
finding is twofold. Firstly, Leicht, Holme and Newman
\cite{Newman2006} have recently proposed a new index of node
similarity, which is actually a variant of the well-known Katz index
\cite{Katz1953}. The numerical tests \cite{Newman2006} showed that
their index best reproduces the known correlations between nodes
when the parameter is very close to its maximal value that
guarantees the convergence of power series expansion. Although their
work and the current work originate from different motivations and
use different testing methods, the results are surprisingly
coincident. Despite the insufficiency of empirical studies and the
lack of analytical insights, this finding should be of theoretical
interests. Secondly, $\varepsilon_{\texttt{max}}$ is equal to the
inverse of the maximum eigenvalue of $\textbf{S}$,
$\lambda_{\texttt{max}}^{-1}$. Therefore, it is easy to determine
$\varepsilon_{\texttt{max}}$ since fast algorithms on calculating
$\lambda_{\texttt{max}}$ for a given matrix is well developed (see,
for example, the \emph{power iteration method} in Ref.
\cite{Matrix}). When dealing with an unknown system, we can first
calculate $\lambda_{\texttt{max}}$, and then concentrate the search
of $\varepsilon_{\texttt{opt}}$ on the area around
$\lambda_{\texttt{max}}^{-1}$, which can save computations in real
applications.

Very recently, a fresh issue is raised to physics community, that
is, how to predict missing links of complex networks
\cite{Clauset2008,Redner2008}. The fundamental problem is to
determine the proximities, or say similarities, between node pairs
\cite{Liben-Nowell,Zhou2009}. The similarity index presented here is
not only an extension of the Pearson correlation coefficient in
rating systems, but also easy to be extended to quantify the
structural similarity of node pair in general networks based on any
locally defined similarity indices. We believe this self-consistent
definition of similarity (see Eq. (3)) can successfully find its
applications in link prediction problem.

We acknowledge \emph{GroupLens Research Group} for \emph{MovieLens}
data (http://www.grouplens.org). This work is supported by the 973
Project 2006CB705500, and the National Natural Science Foundation of
China under Grant Nos. 60744003 and 10635040. T.Z. and J.-G.L.
acknowledge the support from SBF (Switzerland) for financial support
through project C05.0148 (Physics of Risk), and the Swiss National
Science Foundation (205120-113842).

\end{document}